\documentclass[preprint,review,12pt]{elsarticle}

\usepackage{graphicx}
\usepackage{subfig}
\usepackage{soul}
\usepackage{float} 
\usepackage{amssymb}
\usepackage[dvipsnames]{xcolor}
 \usepackage{bm}
\usepackage{geometry}
\usepackage{fancyhdr}

\usepackage{lineno}
\usepackage{pdflscape} 
\usepackage{hyperref} 

\journal{Computers and Geotechnics}

\begin{document}

\begin{frontmatter}


\title{Numerical computation of stress-permeability relationships of fracture networks in a shale rock}



\author{Rafael March, David Egya, Christine Maier, Andreas Busch, Florian Doster}
\address{Institute of Geoenergy Engineering, Heriot-Watt University, Edinburgh, EH14 4AS, UK.}
\address{Correspondence email address: f.doster@hw.ac.uk.}



\begin{abstract}
We present stress-sensitive permeability relationships for two-dimensional fracture networks in the Opalinus Clay from the Mont Terri underground rock laboratory. These relationships may be used as a proxy for fracture network permeability in numerical models that resolve large spatial scales and are used in a variety of geoenergy applications involving flow in shaly rocks. To obtain these relationships we present a numerical procedure that uses experimentally determined stress-permeability relationships to numerically compute the effective permeability of the network. The material discontinuities stemming from the fractures are treated by a simple contact-interaction algorithm that accounts for normal interaction between fracture walls, allowing us to calculate the permeability of a fracture network under different stress conditions.
We apply the procedure to four fracture networks digitized from two galleries of the Mont Terri rock laboratory. These fracture networks are mapped from the damage zone of the Main Fault that intersects the Opalinus Clay. The networks show a maximum variation of four orders of magnitude when stress ranges from 1 MPa to 20 MPa. Our numerical procedure not only establishes representative stress-permeability relationships for a fractured rock mass under stress, but also provides a proxy for fracture network permeability for simulation in fractured formations.
\end{abstract}


\end{frontmatter}


\newcommand{\REF}[1]{\textcolor{red}{[REF] #1}} 
\newcommand{\TODO}[1]{\textcolor{red}{[TODO] #1}}
\newcommand{\UPDATED}[1]{\textcolor{green}{[UPDATED] #1}}


\section{Introduction}\label{S:1}
Flow in fractured rock masses with low matrix permeability is ubiquitous in the context of geoenergy applications. Some of these applications include extraction of hot water from geothermal reservoirs, risk assessment of CO$_2$ leakage through fractured caprocks during CO$_2$ storage operations, and modelling of fluid flow in fractured hydrocarbon reservoirs. In the geothermal energy context, Enhanced Geothermal Systems (EGS) \citep{Breede2013, Held2014} allow us to extract hot water from low-permeability reservoirs by injecting fluid into the subsurface in order to cause pre-existing fractures to re-open or to create a fracture network that allows for economic energy extraction. In carbon storage operations, fractures might be either present in the storage reservoir (e.g. naturally fractured reservoirs) or in the caprock which must prevent CO$_2$ to flow upwards to the surface \citep{March2018}. In hydrocarbon production from fractured reservoirs, fractures often play a key role in the mobility of the different fluids that exist in the reservoir \citep{narr2006naturally}. Hence, a detailed characterization of fluid flow in fracture networks is of utmost importance in the evaluation of the feasibility of many geoenergy-related projects.

Subsurface geological formations are under stress because of rock mass overburden and the larger tectonic environment. Due to the mechanical response of fractures to stress, flow in the fractures is not only controlled by the intrinsic fracture hydraulic apertures, but is also sensitive to in-situ stress conditions. The far-field stresses will generally act to close fractures, while the flowing fluid pressure acts to open them, leading to the concept of an effective stress that controls the fracture network permeability. The permeability is also affected by the roughness of the fracture planes, which affects the intrinsic fracture aperture, i.e. the aperture of the fracture at unloaded conditions. Hence, any modelling effort to estimate the permeability of fracture networks for geoenergy applications require the treatment of these aspects.

There is a major challenge when modelling fluid flow in fractured rocks. Like everything else in porous media, fractures are multi-scale features in the rocks. It is generally impossible to explicitly represent the roughness of the fracture walls and the actual fracture locations in large-scale simulations, since these properties are unknown for actual fracture networks in the subsurface and this would lead to prohibitive computational time. Typically, some method of upscaling is required to translate the stress acting on the fractures of a fracture network to a fracture network stress-permeability relationship that can be used in large-scale models. Constitutive models are generally used for single-fracture stress-aperture relationships. These models come either from numerical simulations that resolve the fracture roughness \citep{McDermott2006, Kubeyev2020} or from experiments that subject a rock mass with a single fracture under different levels of stress and measure the fracture permeability \cite{Barton1980, Zhang2013}. The single-fracture aperture models can then be used in a model with fractures represented as lines (in 2D) or planes (in 3D) for fracture network permeability upscaling.

This paper presents a numerical procedure developed to estimate the stress-permeability relationship in a low-permeability fractured formation. It is applied to calculate effective permeabilities for sections of the Main Fault of the Opalinus Clay in the Mont Terri rock laboratory, in Switzerland \cite{Bossart2017}. The procedure involves linking experimentally determined fracture apertures \cite{Zhang2013} with a linear-elastic mechanical model \cite{beirao2013basic, klemetsdal2016virtual, andersen2017virtual} augmented by a contact interaction algorithm to account for the effect of applied stress on fracture apertures and permeability. The rock matrix is considered impermeable in the caprock so that flow takes place only through the fractures. The resulting model is then used with a Discrete Fracture Network (DFN) for flow-based computation and upscaling of effective permeability. Field test data from the Mont Terri underground laboratory serves as a comparison to the numerical estimates from our computational tools. Our numerical results capture the order of magnitude of change in permeability due to the associated fractures around the Main Fault damage zone. The upscaled permeability results can serve as a proxy for stress-sensitive permeability of fracture networks for modelling single-phase fluid-flow in large-scale numerical simulations in low-permeability fractured formations for geoenergy applications. 

The remaining sections of this paper are organised as follows. First, we present background to our research methodology and the numerical workflow. We also outline the coupled hydro-mechanical model concept and the underlying equations and assumptions. Then, we present numerical results of upscaled stress sensitive permeability of four fracture networks located in two galleries of the Mont Terri rock laboratory. Finally, we derive relevant conclusions on the basis of these results. 

\section{Background and Methodology}\label{S:3}

This section provides background to the numerical framework implemented in the open-source MATLAB Reservoir Simulation Toolbox (MRST) \cite{Lie2019} for the upscaling of fracture network permeability under stress conditions. It also introduces fracture networks mapped from the Main Fault of the Mont Terri rock laboratory. The procedure is applied to these fracture networks for the evaluation of stress-permeability relationships. 

\subsection{Background}

The Mont Terri Project is an underground rock laboratory that aims to investigate and analyse the hydrogeological, geochemical and rock mechanical properties of argillaceous formations for geological disposal of radioactive waste \cite{Bossart2017}. In the laboratory, experiments are dedicated to investigate the properties of the kaolinite-bearing Opalinus Clay Formation (OPA), that has been selected as suitable candidate for long-term underground geological storage. The laboratory is intersected by a major thrust fault, the ``Main Fault", which is ~0.9 to 3 m thick. The galleries drilled across the Main Fault allow the researchers to measure its hydromechanical characteristics, such as fault permeability and fault slip \cite{Laurich2017}. The field scale experiments are particularly relevant to the understanding of fluid flow in fault damage zones, since they provide in-situ measurements of fault permeability.

The Mont Terri Project is an underground rock laboratory that aims to investigate and analyse the geophysical, hydrogeological, geochemical and rock mechanical properties of the Opalinus Clay Formation (OPA) for geological disposal of radioactive waste \cite{Bossart2017}. There are numerous laboratory studies on various properties, making OPA one of the best studied mudrocks formations in the world. The laboratory is intersected by a major thrust fault, the “Main Fault", which is ~1.5 to 4 m thick within the sections along strike that have been penetrated by galleries. These galleries allow the researchers to measure its hydromechanical characteristics, such as fault permeability and fault slip \cite{Laurich2017}. The field scale experiments are particularly relevant to the understanding of fluid flow in fault damage zones, since they provide in-situ measurements of fault permeability. Various field activities have recently taken place to test the frictional and hydrological properties of the Main Fault in the context of CO2 leakage from storage sites and induced seismicity \cite{Cappa2018, Rutqvist2016, Zappone2020}.

\begin{figure}[H]
\centering
\includegraphics[width=0.8\textwidth]{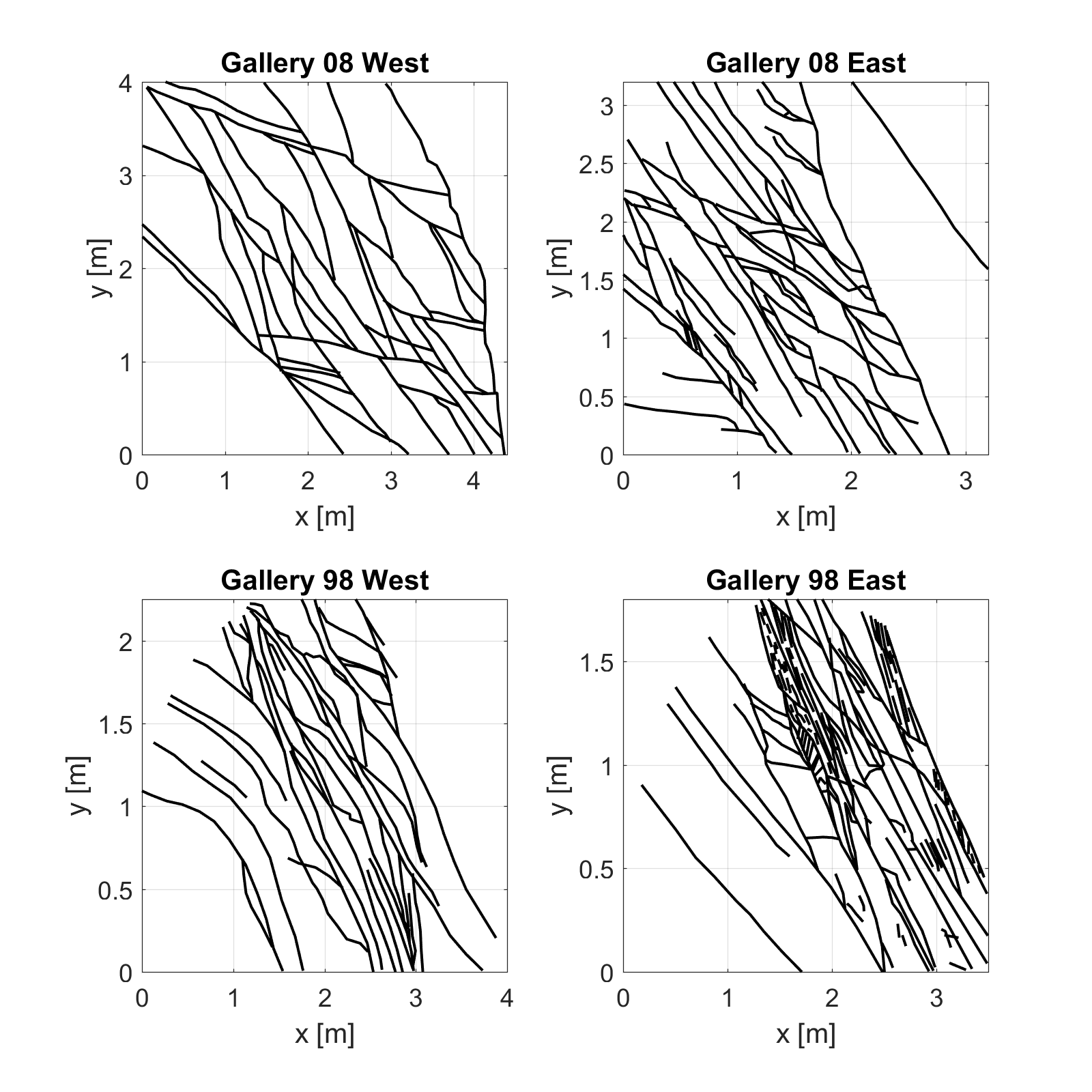}
\caption{Four fracture networks mapped from the outcrops of the Main Fault situated in two Mont Terri galleries, Gallery 08 and Gallery 98. Top row shows the west (left) and east (right) view of the Gallery 08 outcrop. The bottom row shows the west (left) and east (right) view of the Gallery 98 outcrop.}
\label{fig:galleries}
\end{figure}

Four windows present in Gallery 08 and Gallery 98 of the rock laboratory provide us with a view of the structure of small parts of the fracture network of the fault's damage zone. Figure \ref{fig:galleries} shows digitised fracture networks that have been mapped by \cite{Jaeggi2017}. We denote each fracture network according to the respective galleries: Ga08W and Ga08E are the west and east views of the fracture network mapped in Gallery 08 and Ga98W and Ga98E are the west and east views of the fracture network mapped in Gallery 98. East and west windows of each of the two galleries are approximately 4m apart and the two galleries have a distance of about 40-50m along strike of the fault. The fracture networks in the four windows show similar structures, with an apparently well-connected system of fractures and window sizes ranging from 1.8m to 4m in the vertical direction and from 3m to 4m in the horizontal direction.

The windows presented in Figure \ref{fig:galleries} show just a small part of the fracture network of the Main Fault's damage zone. It is still unclear how well connected these fracture networks are between galleries 08 and 98 or what the extent of these networks parallel to the fault plane is. However, the fracture networks give an example for the topology and structure of fault-related fracture networks in shale formations and help us understanding flow behaviour in fractured low-permeability formations.

\subsection{Methodology}

\begin{figure}[H]
\centering
\includegraphics[width=0.9\textwidth]{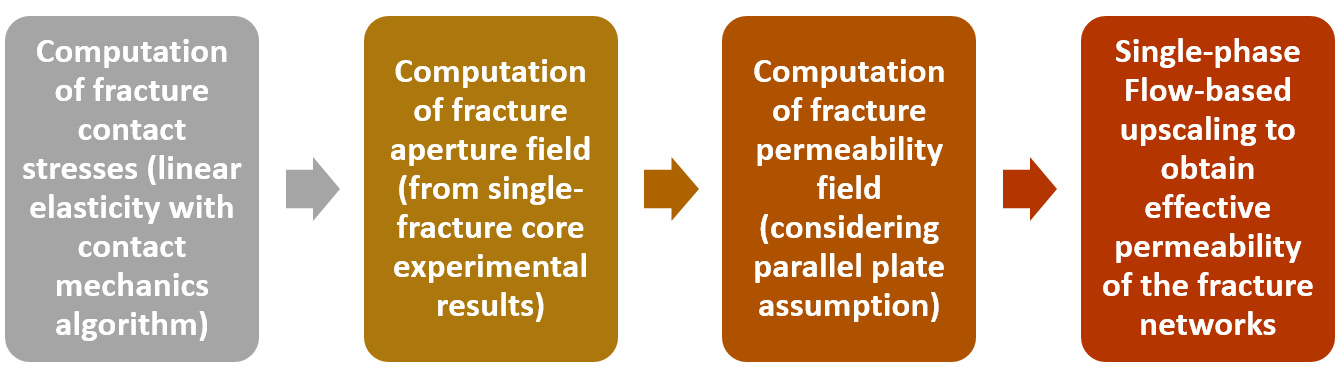}
\caption{Workflow for estimation of the stress-sensitive permeability of fracture networks.}
\label{fig:workflow}
\end{figure}

This section presents a workflow to compute the permeability of fracture networks under different in-situ stress conditions. We apply stresses to an augmented rock mass around the fracture network, compute the contact stresses in the individual fractures, compute the aperture of each fracture segment and then apply flow-based upscaling to compute the effective permeability of the fracture network. The workflow is similar to the one applied in Bisdom et al. (2016) \citep{Bisdom2016}, but it uses fractured core experimental data instead of the empirical model of Barton and Bandis \citep{Barton1980}. This change is important to capture the correct order of magnitude in stress-permeability models for fracture networks in shale formations, as the single-fracture stress-aperture model plays a key role in the final upscaled results for the whole network. Moreover, the model of Barton and Bandis has been developed for granites with apertures on the order of millimeter and its applicability to shales is questionable. Figure \ref{fig:workflow} presents an outline of the workflow to compute the effective permeability for a fracture network.

We start by applying stress boundary conditions to a domain containing the fracture networks of Figure \ref{fig:galleries} (Section \ref{sec:mech}). We consider a buffer zone around the fracture network to avoid boundary effects on the deformation of the fracture planes. Through a simplified contact mechanics algorithm, we process the results to compute contact stresses on each individual fracture, as a result of the applied stresses. The contact stress is the normal stress experienced by the fracture walls and will serve to compute the aperture of each fracture segment. The contact stress acting on each fracture is used to compute an aperture using and experimental model (Section \ref{sec:singlefracap}). The experimental model allows us to reproduce the correct single-fracture stress-aperture behaviour for the Opalinus Clay. Finally, we use the fracture networks and the computed apertures to perform single-phase flow-based upscaling (Section \ref{sec:keffupscaling}). A flow field is computed in the fracture network, disregarding matrix flow. The upscaled permeability results allow us to understand the stress-permeability behaviour of the whole network. 

\subsubsection{Mechanics Model}
\label{sec:mech}

We consider the rock mass around the fractures an isotropic material and solve the equations that model linear elasticity with infinitesimal displacements. The physical model is mathematically described by the quasi-static momentum equations:
\begin{equation}
     \nabla \cdot \bm{\sigma} = \bm{F},
    \label{eq: elasticity1}
\end{equation}
where $\sigma$ is the symmetric Cauchy stress tensor and $\bm{F}$ a body force. The body force $\bm{F}$ is taken as zero in this work. For isotropic materials, the Cauchy stress tensor is given by:
\begin{equation}
     \bm{\sigma } = \bm{C:\varepsilon},
    \label{eq: elasticity2}
\end{equation} 
where strain $\bm{\varepsilon}$ is the symmetric part of the displacement gradient, $\bm{\varepsilon}= \frac{1}{2}(\nabla + \nabla^{T})\bm{u}$, and $\bm{C}$ is the stiffness tensor. The components of the stiffness tensor for an isotropic material are given by:
\begin{equation}
     C_{ijkl} = \lambda\delta_{ij}\delta_{kl} + \mu(\delta_{ik}\delta_{jl}+\delta_{il}\delta_{jk}),
    \label{eq: elasticitye}
\end{equation}
where $\lambda$ and $\mu$ are the first and second Lame parameters which are related to the Poisson's ratio ($\nu$) and Young's modulus ($E$) by:
\begin{eqnarray}
    \lambda &=& \frac{E\nu}{(1+\nu)(1-2\nu)}, \\
    \mu &=& \frac{E}{2(1+\nu)}.
\end{eqnarray}
We take $\nu = 0.35$ and $E = 2.5$ GPa as representative values for shale rocks \citep{Pearson2003}.

\begin{figure}[H]
\centering
\includegraphics[width=0.4\textwidth]{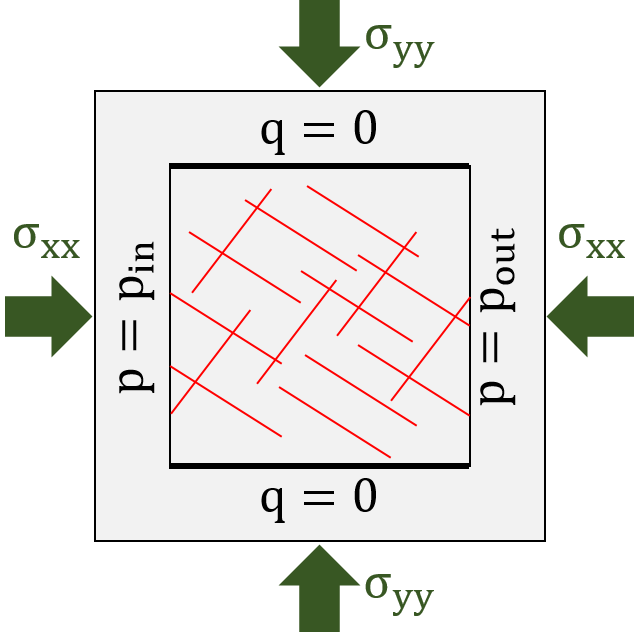}
\caption{Mechanics and flow simulation boundary conditions. The area outside the flow boundary is impermeable.}
\label{fig:BC}
\end{figure}

Figure \ref{fig:BC} shows a schematic of the domain and the boundary conditions for the mechanical and the flow problem. We apply constant stress boundary conditions along each boundary edge but their values differ in each direction. We also constrain both displacements at the mid nodes of the mesh of each boundary edge, to prevent translation and rotation of the model ($\bm{u}(\bm{x_m})=\bm{0}$).

Fractures are modelled as material discontinuities that lie within the domain. This means that walls of a fracture are disconnected and can displace independently (displacement discontinuity is allowed across the fractures). This is necessary to capture the impact of stresses on the fracture aperture. However, this also requires the treatment of the fracture walls as a hard contact within the material domain. The contact stress obtained from the solution of the contact mechanics problems. In the contact mechanics problem, we disregard the shear displacement of the fractures. This means that the fractures are only allowed to displace in the direction normal to the fracture plane (Mode I opening). This assumption significantly simplifies the contact mechanics algorithm. It also does not capture local strain deviations caused by the shear displacement of some fractures that could induce local perturbations in the contact stress of other fractures. Other effects such as shear dilation (opening of fracture due to shear) are also not captured with this approach.

\subsubsection{Single-Fracture Stress-Aperture Model}
\label{sec:singlefracap}

We use the experimental data of Zhang (2013) \cite{Zhang2013} as a model for the single fracture stress-permeability in the Opalinus Clay. In Zhang (2013) the author considers a core containing few mostly parallel fractures extracted from the Opalinus Clay and measures permeability to a gas phase under different confining pressures. The measured permeability includes a contribution from the core's matrix and a fracture network that are accounted for by considering the following effective model:
\begin{equation}
k_g = k_m + \frac{F}{12}\left(b-b_c\right)^3,
\label{eq:kgcore}
\end{equation}
where $k_g$ is the measured permeability to gas, $k_m$ is the intact rock matrix permeability, $b$ is the fracture aperture, and $F$ and $b_c$ are fitting parameters. The average fracture aperture $b$ is expressed as:
\begin{equation}
     b = b_m e^{-\alpha \sigma_n^\beta},
\label{eq:aperture}
\end{equation}
where $b_m$ is the unstressed fracture aperture, $\sigma_n$ is the contact stress on the fracture segment, and $\alpha$ and $\beta$ are fitting parameters. The unstressed aperture is also a fitting parameter. A point of note is that this model assumes that fractures may be completely closed if subject to sufficiently high stress ($b \rightarrow 0$ when $\sigma_n \rightarrow \infty$). This assumption differs from other models available in the literature, such as the Barton and Bandis model, which assumes that the maximum closure is less than the initial (unloaded) aperture.

\begin{figure}[H]
\centering
\includegraphics[width=\textwidth]{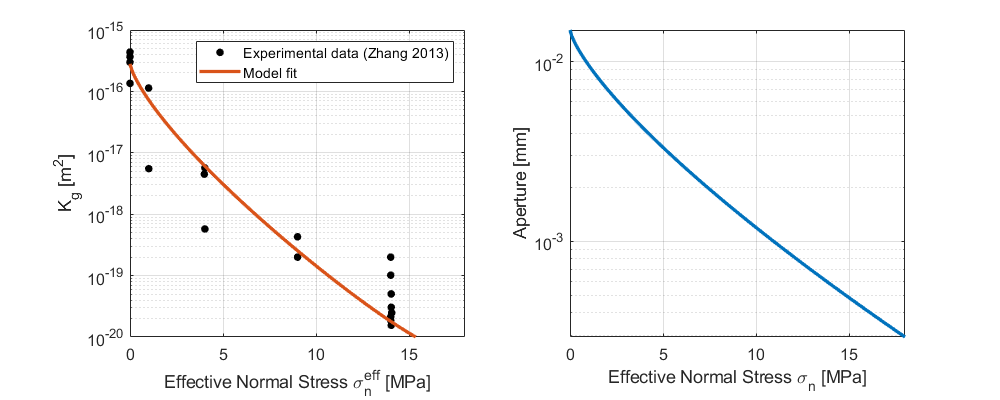}
\caption{Fracture core experimental data and model fit. Left: Permeability to gas phase and fit of model defined by Equation (\ref{eq:kgcore}). Right: Corresponding stress aperture model for a single fracture derived from the permeability curve using Equations (\ref{eq:aperture}) and (\ref{eq:kgcore}).}\label{fig:core}
\end{figure}

Figure \ref{fig:core} shows the model fit to the experimental results of Zhang (2013). We perform manual tuning of the fitting parameters described above to adjust the permeability model given by Equation (\ref{eq:kgcore}) to the experimental data points (Figure \ref{fig:core}(a)). We set $k_m = 10^{-21}$ $\rm m^2$ as the matrix permeability, as reported in Zhang (2013). The fitting parameters for the final fitted model are $\alpha=0.45$, $\beta=0.75$, $F = 1$, $b_c = 0$ and an initial aperture of $b_m = 0.015$ mm. Figure \ref{fig:core}(b) shows the stress-aperture model resulting from the fitting procedure (Equation (\ref{eq:aperture})).
 
\subsubsection{Flow-Based Permeability Upscaling}
\label{sec:keffupscaling}

The aperture/permeability model allows us to evaluate the aperture and the permeability of the fractures based on the contact stress acting on them. These properties are then fed into a flow simulator that computes the fracture transmissibilities and allows us to obtain a flow field for the whole network given a fluid pressure gradient. We consider Darcy's model to obtain an effective network permeability, given the volumetric flow rate $q$ and the applied pressure gradient $\Delta P$:
\begin{equation}
    k_{eff} = \frac{q \mu}{A}\frac{L}{\Delta P},
    \label{eqn:darcy}
\end{equation}
where $\mu$ is the fluid viscosity, taken as $\mu = 1$ cP. The cross-sectional area $A$ is the length of the domain edge perpendicular to the flow direction and $L$ is the length of the domain edge parallel to the flow direction. Note that we consider the matrix in the upscaling procedure: the cross-sectional area is not only given by the fracture apertures. This is important if one wants to apply the final obtained effective permeability models in larger scale coarse grid simulations. 

\subsubsection{Numerical Implementation}

\begin{figure}[H]
\centering
\includegraphics[width=0.8\textwidth]{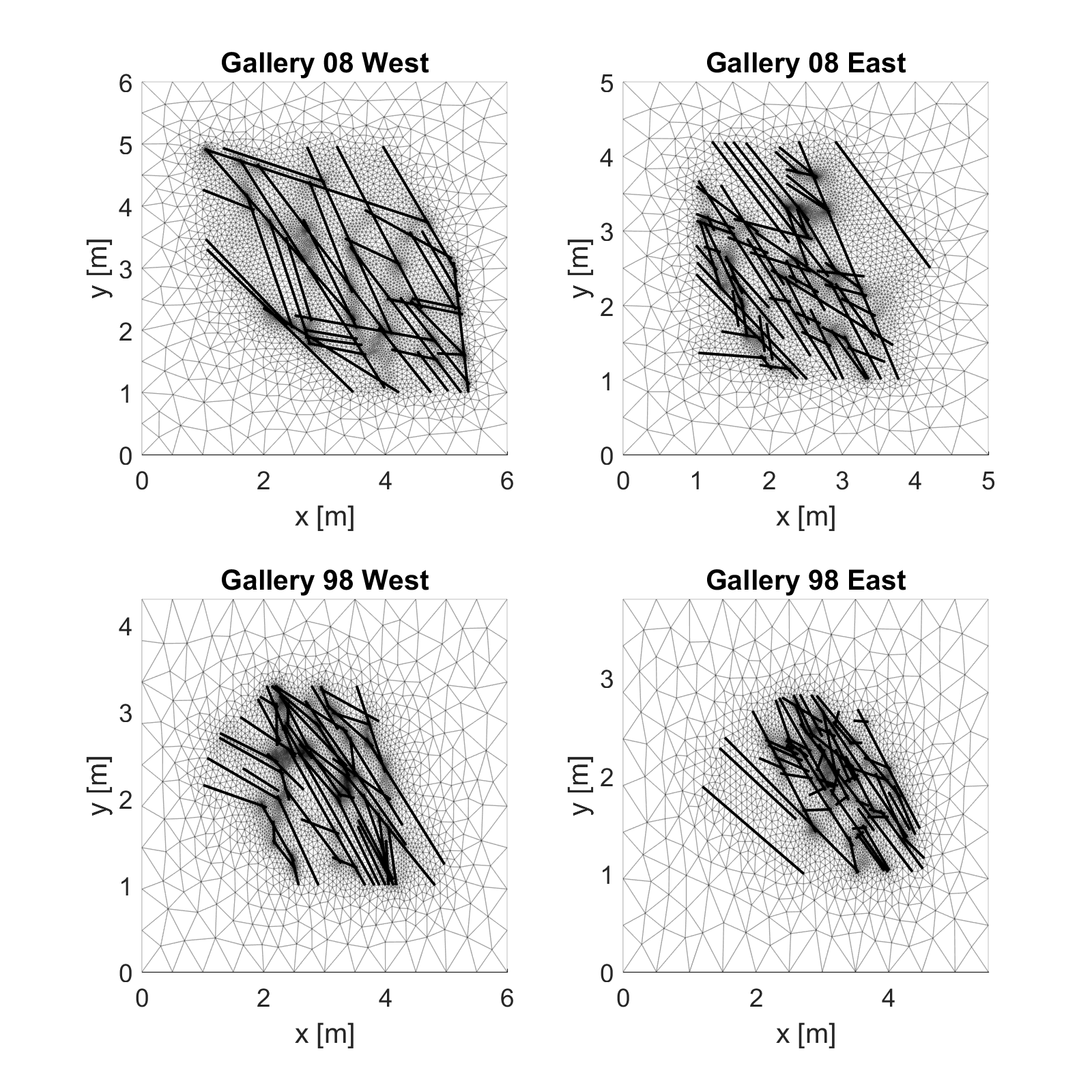}
\caption{Computational triangular mesh of the four fracture networks Ga08W, Ga08E, Ga98W, and Ga98E. The material buffer surrounding the fracture networks eliminates boundary effects for geomechanical computation. This buffer is removed for flow-based upscaling. }  
\label{fig:meshes}
\end{figure}

Our workflow is implemented in MRST, the Matlab Reservoir Simulation Toolbox \cite{Lie2019}. We discretise the two-dimensional domains using a finite element mesh. We use a triangular mesh that conforms to the fracture network. The curved fractures are approximated by a series of linear segments in order to represent the whole network as a set of linear fractures with two end points. Fractures with high tortuosity are difficult to be represented by only two end points and hence are split into multiple fracture segments. The topology and approximate geometry of the fracture networks were preserved to respect the correct fracture network connectivity and aperture distribution. Figure \ref{fig:meshes} shows the triangular meshes generated around the linearized fracture networks. We duplicate each node lying on a fracture edge to obtain a `master' and a `slave' node in order to allow discontinuous displacements across the fracture edges. Intersections between two fractures are split into four nodes. End points are not duplicated, respecting the approximate `seam' representation of a fracture in a rock mass. A mesh with consistent node/edge duplication is an input to our numerical simulator. We have used FractureNetwork2d class within the open-source package PorePy  \citep{PorePy} to generate such meshes. 

For solving the hydromechanical problem, each fracture segment is also associated with a corresponding `virtual cell'. This cell is a one-dimensional element along the fracture segment that has a pore volume in the computational domain, but has no volume in the geometrical mesh. The pore volume of the fractures is equal to the aperture times the segment length. We approximate the flow in the fractures as Darcy flow with a unity porosity and a permeability given by the cubic law with the hydraulic aperture from Equation \ref{eq:kgcore}. Note that the aperture and the permeability of each fracture segment comes from the aperture/permeability model (Equation \ref{eq:aperture}), where the normal stress comes from the contact pressure in the fracture segment.

We consider the virtual element method with linear triangular elements. The virtual element method is implemented in MRST for linear elasticity problems and has been applied to a series of problems related to computational geosciences \citep{andersen2017virtual}, \citep{fumagalli2019dual}. For the fluid flow in virtual cells, we consider a finite volume discretization with the standard two-point flux approximation to relate the flux across connections between virtual cells to the cell pressures. We compute the transmissibilities of each virtual connection based on the individual transmissibilities of the fracture segments, which in turn are calculated based on the segments apertures and permeabilities. For the intersections, we use the star-delta approximation to compute the transmissibility without the need of a small volume located at the intersection \citep{KarimiFard:2004}. The code for the presented workflow is available for download in \citep{mrstfracpackmodule}.

\section{Numerical Simulation Results} \label{S:4}

In this section, we present the numerically computed results of the stress-permeability relationship for the digitised Mont Terri fracture networks (Ga08W, Ga08E, Ga98W, and Ga98E). We start by comparing the computed fracture network effective permeability using the workflow detailed in the previous sections to the in-situ permeability tests carried out by Kneuker et al (2017) \citep{Kneuker2017}. Then we vary the range of the stress boundary conditions to obtain a stress - fracture network permeability relationship that can be used in large scale simulations. 

\subsection{Comparison to In-situ Permeability Measurements}
\begin{figure}[H]
\centering
\includegraphics[width=0.4\textwidth]{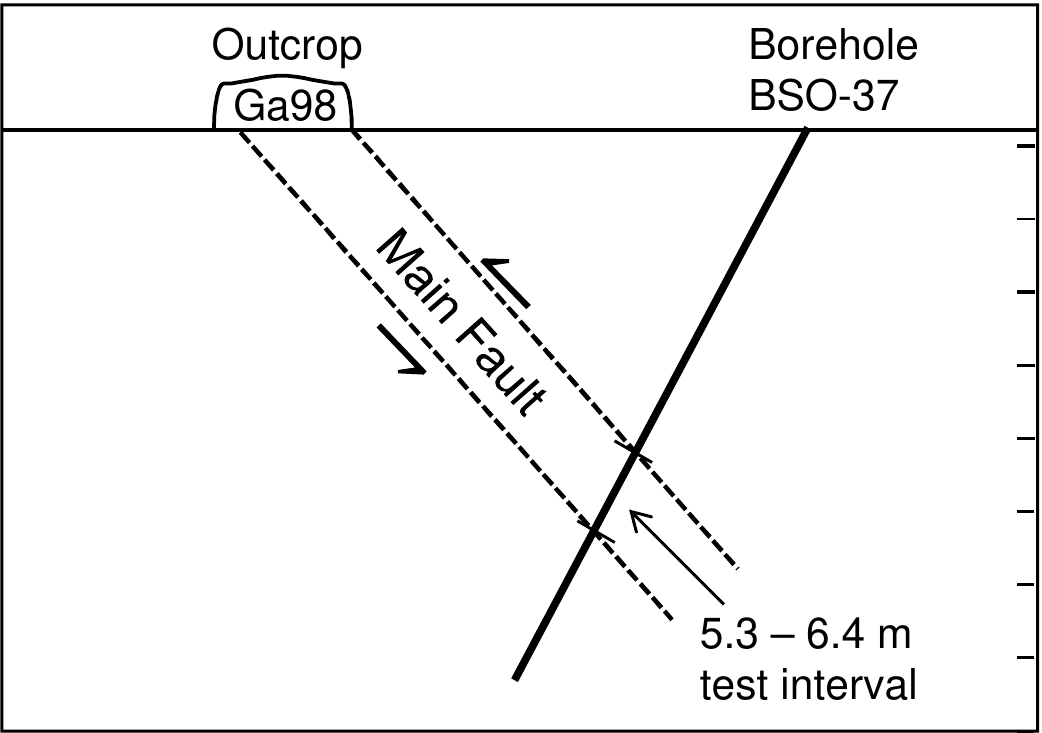}
\caption{Schematic illustration of borehole BSO-37 in relation to the Main Fault and fracture network Ga98. (After \citet{Kneuker2017})}\label{fig:BSO-37}
\end{figure}

In Kneuker et al (2017), Nitrogen was injected at a pressure of $P_{inj} = 0.2$ MPa at different depths along the borehole named BSO-37. The injection was carried out in form of a pulse test into the packed-off interval, followed by monitoring of pressure evolution. Figure \ref{fig:BSO-37} shows a schematic of the intersection of the Main Fault with Gallery 98 and the location of the borehole. The measured permeability of the Main Fault is between $1.0 \times 10^{-19}$ $\rm m^{2}$ and $2.0 \times 10^{-18}$ $\rm m^{2}$. The numerical values are shown in Table \ref{tab:perm}.

\begin{table}[H]
\centering
\begin{tabular}{ccc}
Middle of Interval {[}m{]} & Permeability {[}m$^2${]} \\
5.3                        & 1.0E-19                 \\
5.67                       & 2.0E-18                 \\
6.1                        & 1.0E-18                 \\
6.4                        & 7.0E-19                
\end{tabular}
\caption{Measured values of permeability of the Main Fault. (Extracted from \citet{Kneuker2017})}
\label{tab:perm}
\end{table}

\begin{figure}[H]
\centering
\includegraphics[width=\textwidth]{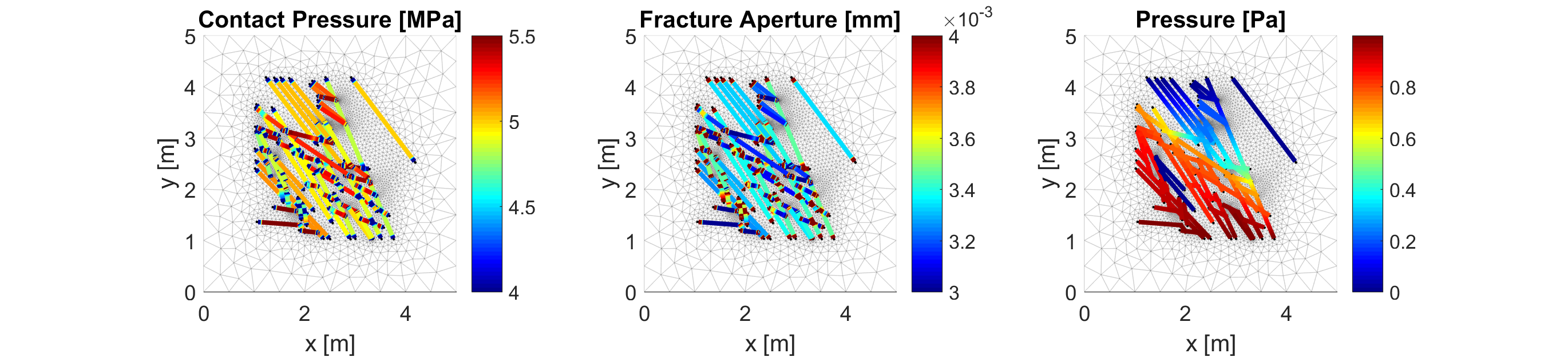}
\caption{Scalar fields in the fracture network Ga08E for comparison with in-situ permeability measurements: contact stresses (left), hydraulic aperture (middle) and pressure(right).}\label{fig:fields}
\end{figure}

Stress boundary conditions of $\sigma_{yy} = 5.5$ MPa and $\sigma_{xx} = 4.7$ MPa are applied in order to compute the stress field. Figure \ref{fig:fields} shows the computed contact stress, aperture and permeability fields in the fracture network Ga08E. We observe that the oblique, nearly vertical fractures experience less stress than the horizontal fractures rendering them to be more open and permeable for flow. This also reflects in the resulting higher effective permeability in y direction ($k_{eff}^y = 3.5 \times 10^{-18}$ $\rm m^2$) compared to the upscaled fracture network permeability in x direction ($k_{eff}^x = 3.2 \times 10^{-18}$ $\rm m^2$). The other three fracture network windows show similar values with an effective permeability in x direction of $k_{eff}^x = 2.2 \times 10^{-18}$ $\rm m^2$, $k_{eff}^x = 5.2 \times 10^{-18}$ $\rm m^2$, $k_{eff}^x = 4.7 \times 10^{-18}$ $\rm m^2$, and an effective permeability in the y direction $k_{eff}^y  = 5.7 \times 10^{-18}$ $\rm m^2$, $k_{eff}^y = 8.5 \times 10^{-18}$ $\rm m^2$, and $k_{eff}^y = 6.9 \times 10^{-18}$ $\rm m^2$ for fracture networks Ga08W, Ga98E, and Ga98W, respectively. 

\begin{figure}[H]
\centering
 \includegraphics[width=0.5\textwidth]{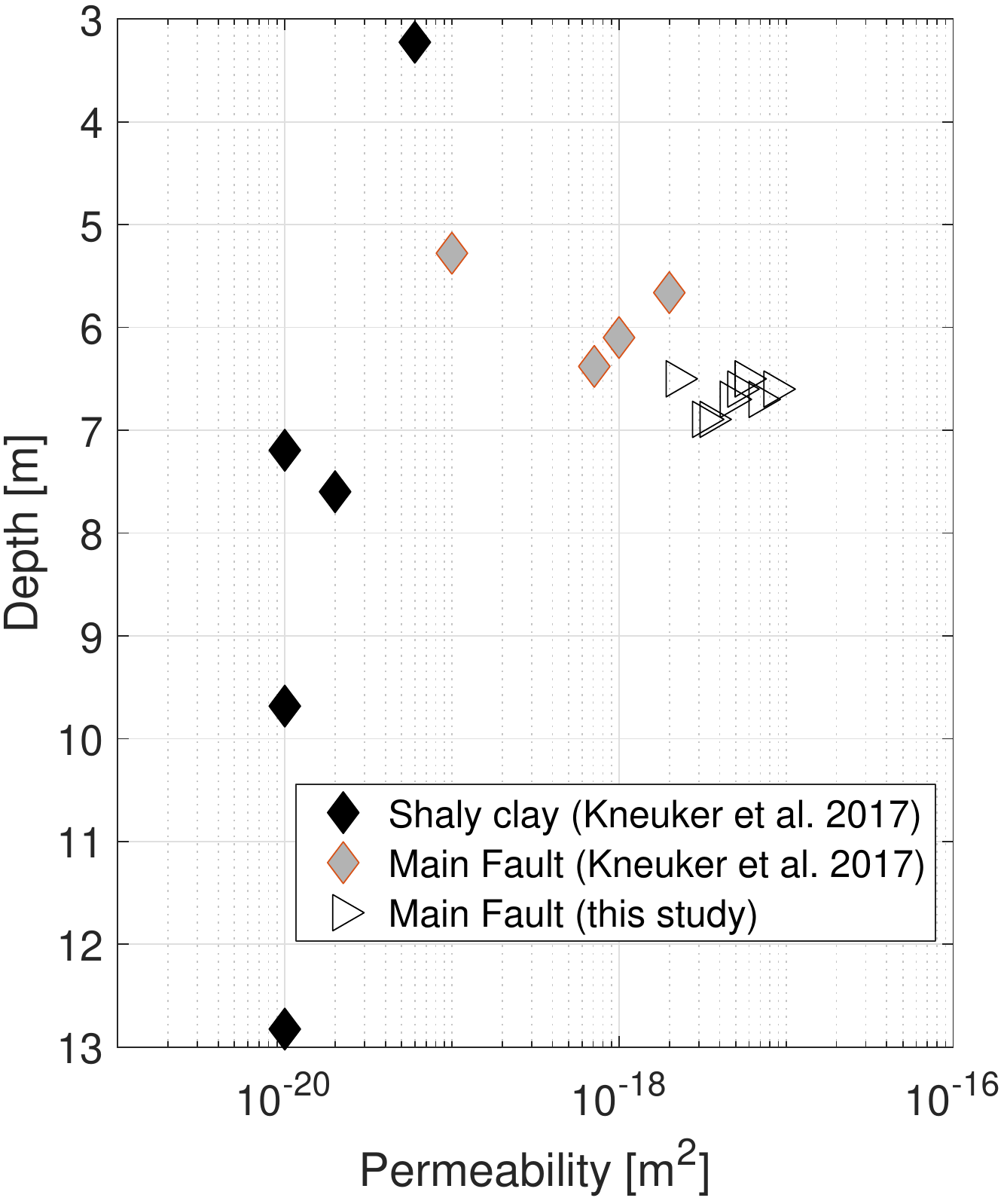}
 \label{fig:kfault_x}
\caption{Permeability of the clay rich shale and main fault in $x$- (a) and $y$-direction (b). Diamonds correspond to the measurements reported in Kneuker et al. 2017 and in Table \ref{tab:perm}. Triangles correspond to the values computed in this study. Fractures increase this base permeability by over $2$ orders of magnitude.} 
\label{fig:kfault}
\end{figure}

 As shown in Figure \ref{fig:kfault}, these results are slightly higher than the in-situ permeabilities reported by \citet{Kneuker2017} and shown in Table \ref{tab:perm}). The difference between the values from our numerical estimations and the in-situ measurements could be attributed to a number of reasons. One obvious reason is the difference in the flow direction for the pulse test and our numerical simulations as illustrated in Figure \ref{fig:BSO-37}. Another is the variation in space of permeability measurements. While the in-situ measurements represents permeability in $3$D space, our numerical estimates are for $2$D space. The difference may be due to the third dimension that is missing in our model and could not be directly mapped from the outcrop. Variation in the area and/or radius of investigation between the pulse test and the flow boundaries in our numerical simulations can also account for the differences between the results, given that spatial distribution and connectivity of fractures vary between the dimensions. This underlines the difficulties in appropriately representing the in-situ flow with a $2D$ model. Finally, the stress aperture relationship derived from the experimental data reported in \citep{Zhang2013} might not be entirely representative because the fractures in the core used in the experiments might have different angles to bedding as the natural fractures. 

Nonetheless, the results (shown in Figure \ref{fig:kfault}) are similar to the in-situ values and are within the order of magnitude in permeability increase caused by the fractures in the damage zone around the Main Fault as reported in other studies collated by \citet{Bossart2017}. The results are also comparable to each other. This is expected given their relatively close distance of separation. The distance between Gallery 08 and Gallery 98 is about $40$ $\rm m$. These results showcase the reliability of our numerical procedure to capture the deformation of and flow in fracture networks associated with damage zones in a manner that is consistent with the in-situ response.

\subsection{Stress - Effective Permeability Relationship}

\begin{figure}[H]
\centering
 \includegraphics[width=\textwidth]{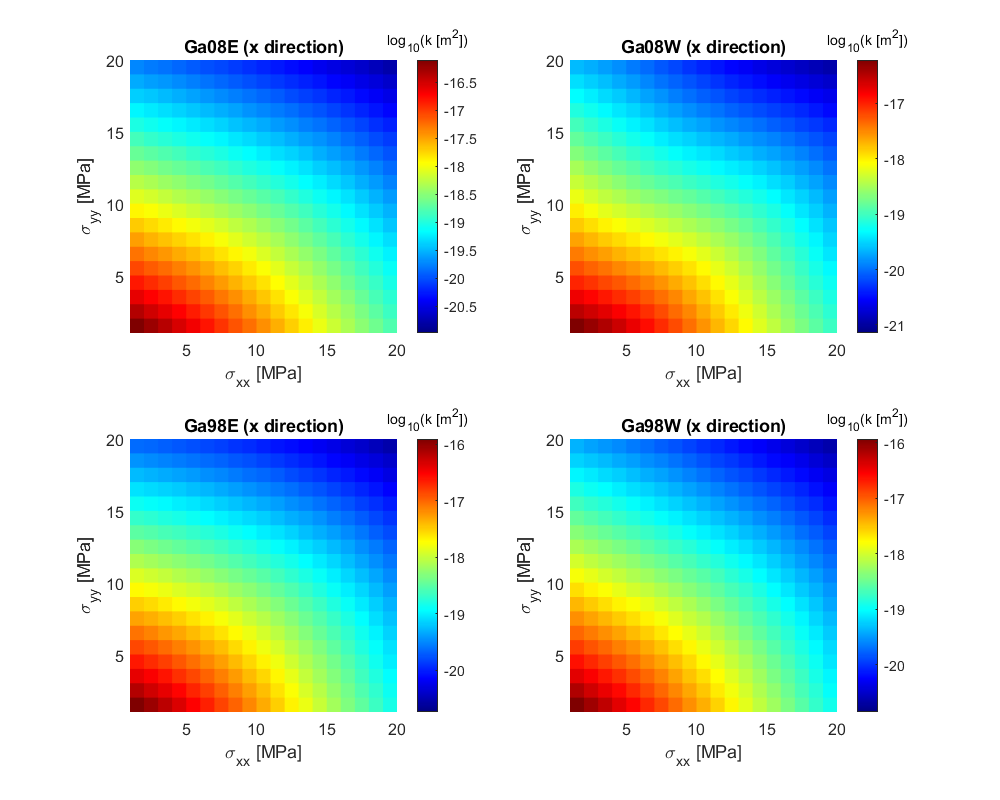}\label{fig:keff_xdir}
\caption{Stress-permeability surface plots for fluid flow in x-direction (a) Ga08E, (b) Ga08W, (c) Ga98E, and (d) Ga98W.}. 
\label{fig:stress-keff-x}
\end{figure}

\begin{figure}[H]
\centering
 \includegraphics[width=\textwidth]{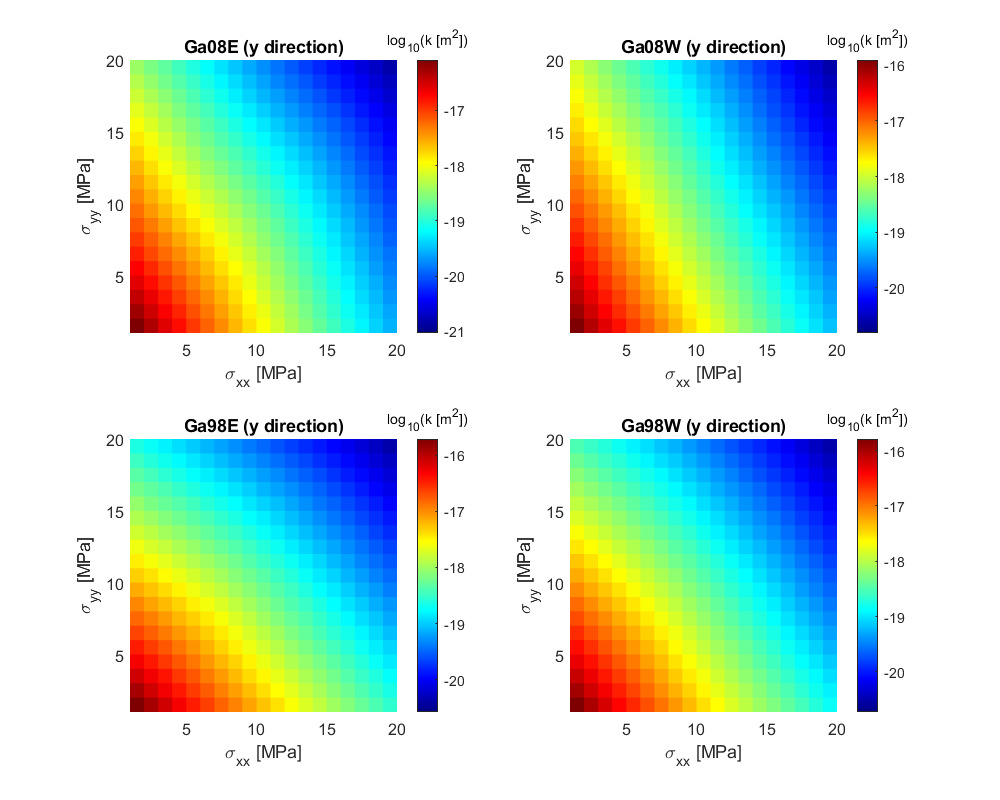}\label{fig:keff_ydir}
\caption{Stress-permeability surface plots for fluid flow in y direction (a) Ga08E, (b) Ga08W, (c) Ga98E, and (d) Ga98W.}. 
\label{fig:stress-keff-y}
\end{figure}

The similarity of our results to the in-situ measurements provides further confidence to use the procedure in predicting the stress - effective permeability relationship for fault associated fracture networks under different stress conditions. For this, we run simulations on each of the fracture networks considering a range of stress boundary conditions ($1.0$ $\rm MPa$ to $20.0$ $\rm MPa$) applied in vertical and horizontal directions. Figures \ref{fig:stress-keff-x} and \ref{fig:stress-keff-y} present the effective permeability in log scale in $x$ and $y$ directions, respectively. Tables containing the values of the surface plots on these figures are presented in \ref{appendix_A}. The results show that upscaled fracture network permeability ranges over five orders of magnitude under different stress conditions. It is highest for low stresses and reduces with increased stress as a consequence of increased normal load on the individual fractures as reported in other studies \cite{cuss2011fracture, Zhang2013, Kneuker2017}. 

The effective upscaled permeability in x direction seems to depend equally on the applied vertical and horizontal stresses resulting in a nearly radial distribution of the permeability values observed in the surface plot (Fig.\ref{fig:stress-keff-x}). The permeability is bound to the lower limit of the intact rock matrix permeability of $k_m = 10^{-21}$ $\rm m^2$ when high horizontal and vertical stresses are applied simultaneously. In case of more heterogeneous stress regime, the computed effective permeability is always above the matrix permeability representing the well connected fracture network. In contrast to very symmetrical upscaled permeability values for horizontal flow as a function of the applied stresses, for the vertical flow we observe a clear elevated dependency on the horizontal stresses. The permeability is decreasing faster with the increase of $\sigma_{xx}$ compared to smaller slopes when the vertical stress, $\sigma_{yy}$ is varied (Fig.\ref{fig:stress-keff-y}). The reason for this lays in the fracture network geometry. All considered fracture networks Ga98W, Ga98E, Ga08W, and Ga08E have certain characteristics in common: The long fractures extending into the y direction are nearly vertical, meaning they will experience more normal stresses induced from $\sigma_{xx}$ than from $\sigma_{yy}$ that is aligned with these fractures in an almost parallel way. At the same time these fractures are the main connectors of the flow boundaries just multiplying the effect of higher stress dependency. On the other hand, the horizontal flow is mainly supported by small fractures intersecting the vertical fractures rendering them to be less influential to the overall flow and thus to the upscaled effective permeability.      

\begin{figure}[H]
    \centering
    \includegraphics[width=0.45\textwidth]{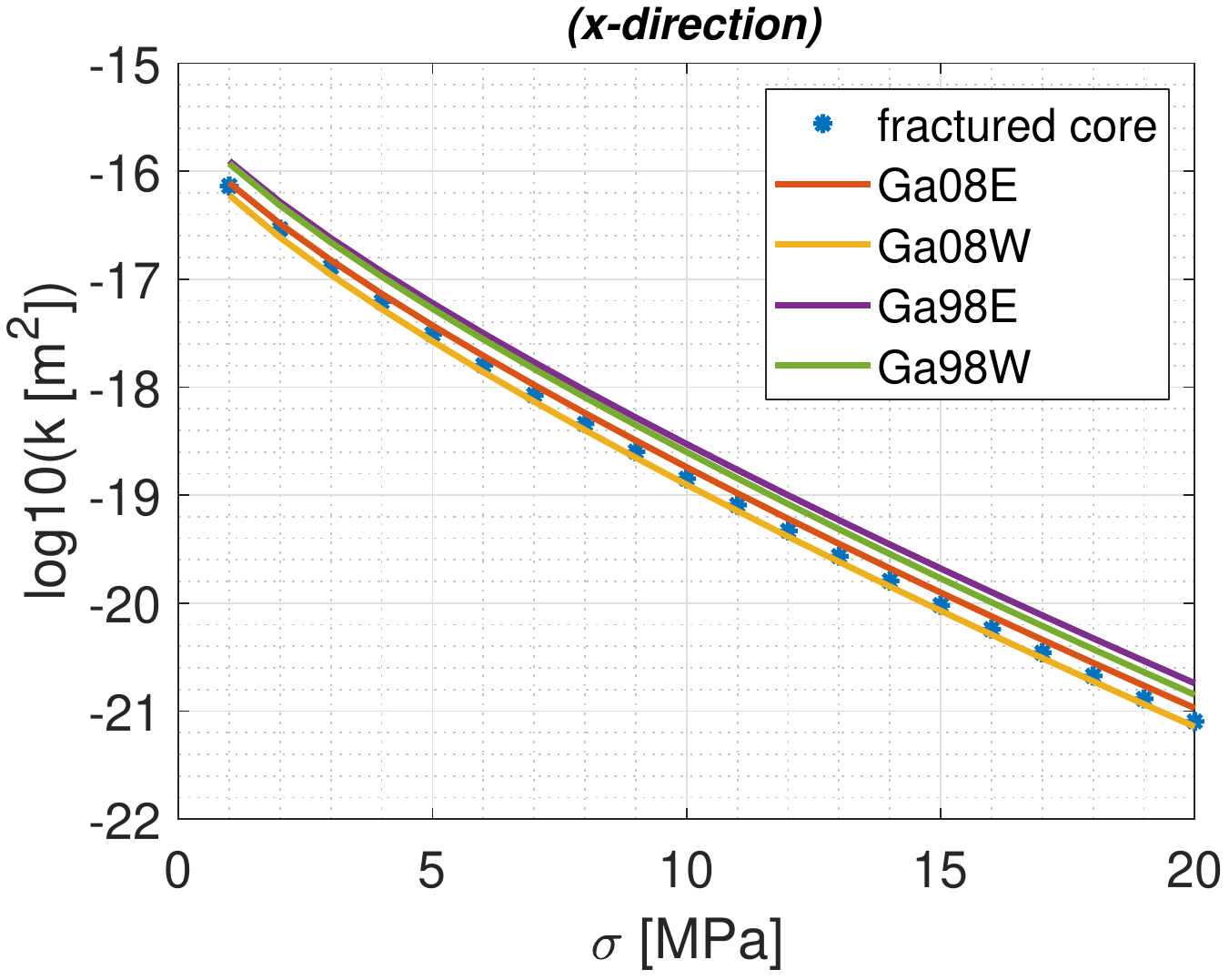}
    \includegraphics[width=0.45\textwidth]{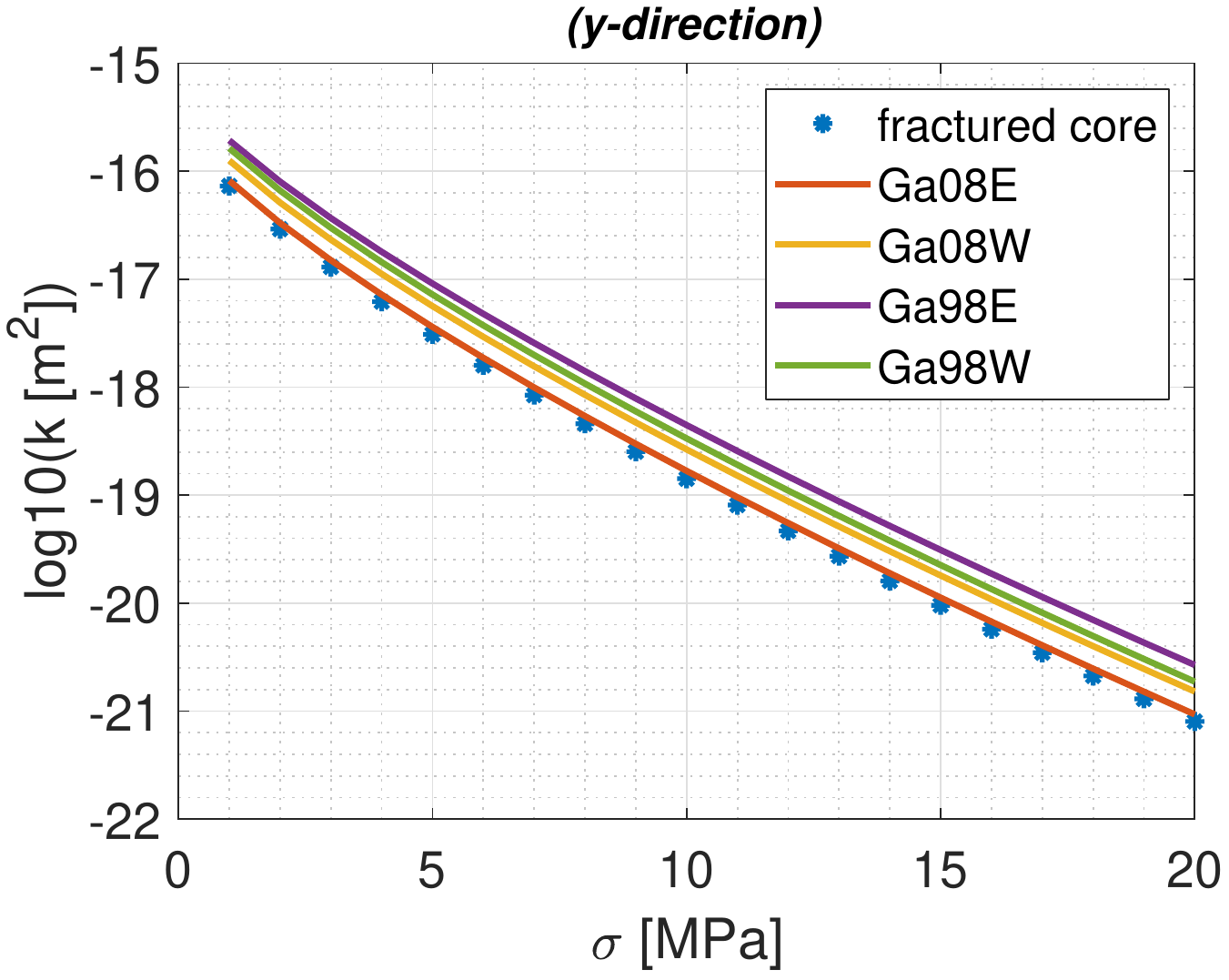}
     \caption{Comparison of stress-sensitive effective permeability for main fault galleries in x and y directions under isotropic stress conditions.} 
    \label{fig:diagonal_group}
\end{figure}

The described effect of fracture orientation density, and connectivity can also be observed in Figure \ref{fig:diagonal_group}. Here, we plot the upscaled effective permeabilities for the four fracture networks in isotropic stress conditions ($\sigma_{xx} = \sigma_{yy}$) against the permeability of the fractured Opalinus Clay sample from Zhang (2013) \cite{Zhang2013}. Under such conditions there is only very little variation in resulting contact stresses on individual fractures. This leads to a very homogeneous aperture and thus permeability field that follows the same slope as the lab measurements. The difference in effective permeability between the individual fracture networks is therefore mainly a product of fracture orientation, density and connectivity. The results show that the permeability in the fracture networks is isotropic to some extent and only slightly higher in y-direction in some cases. As stated above the higher vertical permeability corresponds to the orientation of the larger fractures in the networks, while the shorter fractures strike in x-direction and provide connectivity between fractures in the networks (Fig.\ref{fig:galleries}). 

Note that the effective fracture network permeability appears to be very close if not identical to the measured permeability of the fractured Opalinus clay core in the lab but this is merely a coincidence. If the mapped fracture networks would consist of more fractures that would increase the connectivity, the effective permeability curve would be some factors higher than the lab measurements. On the other hand less fractures and hence poorer connectivity would result in effective permeability even smaller than the laboratory results. Therefore, it is very important to take into account the fracture network rather than just a single fracture permeability when choosing permeability proxies for simulation of flow on coarse scale.  

\section{Conclusions}\label{S:7}
We present a systematic procedure for numerical computation of effective permeability for real fracture networks under different stress conditions obtained from Opalinus clay caprock. The procedure incorporates experimental data from the formation outcrop into a virtual-element-based mechanics solver and discrete-fracture-matrix flow simulation to evaluate the effective permeability of a fault related damage zone. This enable us to establish a stress–permeability relationship that well represents the in-situ conditions and also serve as proxy model for a large scale simulation and formation characterisation for geological CO2 storage. 
\section{Acknowledgement}
We thank Christophe Nussbaum for very valuable discussions dealing about the Mont Terri project in general and the Main Fault in particular. Part of this project has been subsidised through the ERANET Cofund ACT (Project no. 271497), the European Commission, the Research Council of Norway, the Rijksdienst voor Ondernemend Nederland, the Bundesministerium für Wirtschaft und Energie, and the Department of Business, Energy \& Industrial Strategy, UK. This work is also part of a project that has received funding by the European Union’s Horizon 2020 research and innovation programme, under grant agreement number 764531.
\appendix
\section{Table of permeability data in $log_{10}(k[m^2])$ for Figures \ref{fig:stress-keff-x} and \ref{fig:stress-keff-y} surface plots.}\label{appendix_A}

\begin{figure}[H]
\centering
\includegraphics[width=0.8\textwidth]{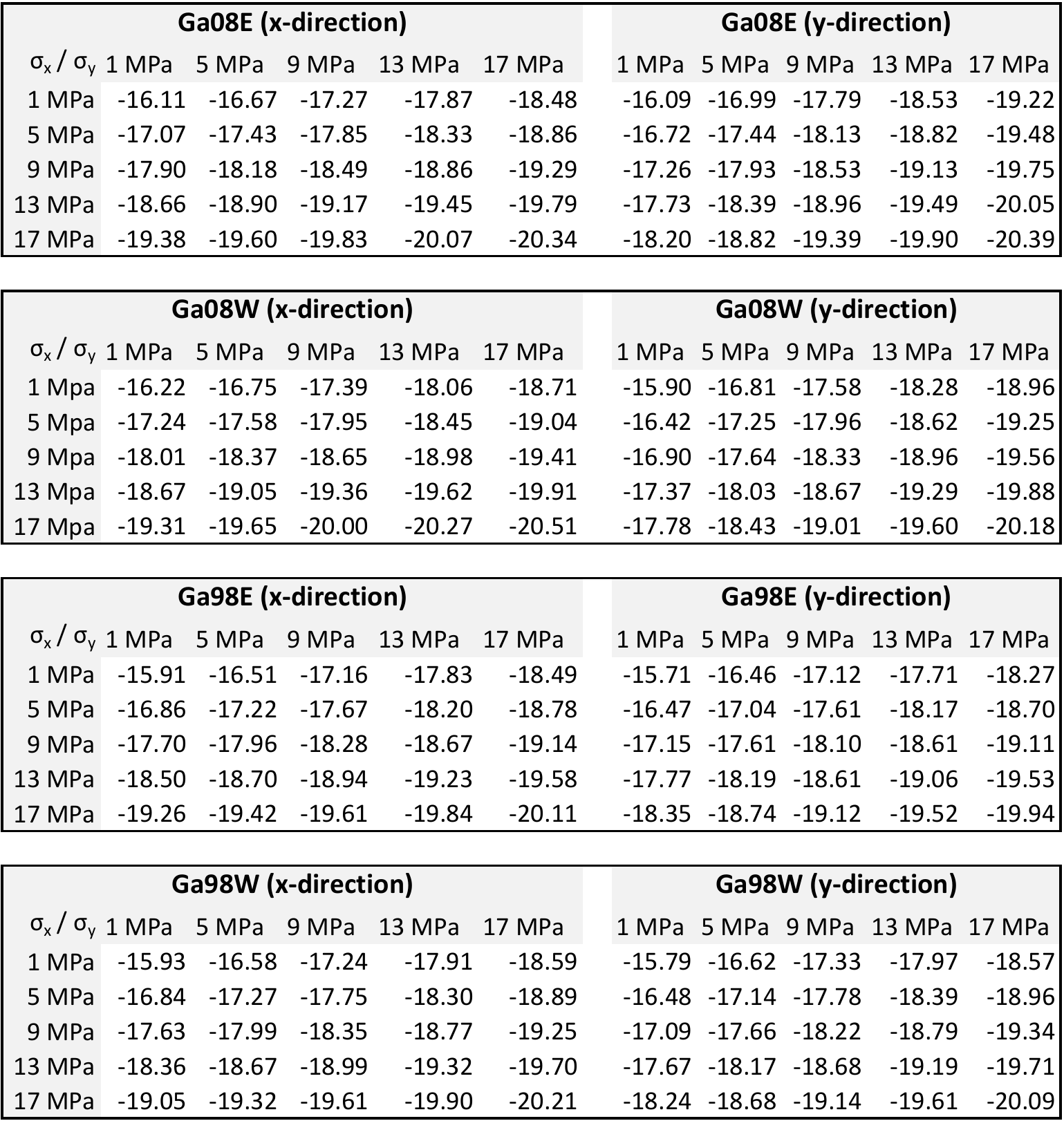}
\end{figure} 




\bibliographystyle{elsarticle-num-names}
\bibliography{mtterri_bib.bib}







\end{document}